# ELECTROMAGNETIC CASCADES AND CASCADE NUCLEOSYNTHESIS IN THE EARLY UNIVERSE

R.J. Protheroe[1], T. Stanev[2] and V.S. Berezinsky[3]
[1]Department of Physics, University of Adelaide, SA 5005, Australia.
[2]Bartol Research Institute, University of Delaware, Newark, DE 19716, U.S.A.
[3]INFN Laboratori Nazionali del Gran Sasso, 67010 Assergi (AQ) Italy.

We describe a calculation of electromagnetic cascading in radiation and matter in the early universe initiated by the decay of massive particles or by some other process. We have used a combination of Monte Carlo and numerical techniques which enables us to use exact cross sections, where known, for all the relevant processes. In cascades initiated after the epoch of big bang nucleosynthesis $\gamma$-rays in the cascades will photodisintegrate $^4$He, producing $^3$He and deuterium. Using the observed $^3$He and deuterium abundances we are able to place constraints on the cascade energy deposition as a function of cosmic time. In the case of the decay of massive primordial particles, we place limits on the density of massive primordial particles as a function of their mean decay time, and on the expected intensity of decay neutrinos.

# 1 Introduction

Electromagnetic cascades in the early universe can be initiated by the decay of massive particles [1, 2], or by their annihilation, by cusp radiation of ordinary cosmic strings [3], by super-massive particles "evaporating" from superconducting cosmic strings [4, 5, 6], by evaporation of primordial black holes, and probably by some other processes. These cascades result in the production of $^3$He and D by disintegration of $^4$He by photons in the cascade, and we shall refer to this as "cascade nucleosynthesis". The epoch of interest for cascade nucleosynthesis has redshift $z \lesssim 10^6$, by which time production of the light elements D, He and Li by big bang nucleosynthesis has taken place. At this epoch, electron-positron pairs are no longer in equilibrium, and black body photons, $\gamma_{\rm bb}$, constitute the densest target for electromagnetic cascade development.

A cascade is initiated by a high energy photon or electron, and develops rapidly in the radiation field mainly by photon-photon pair production and inverse Compton scattering:

$$e + \gamma_{\rm bb} \to e' + \gamma', \quad \gamma + \gamma_{\rm bb} \to e^+ + e^-. \tag{1}$$

Such electron–photon cascading through radiation fields involving photon–photon pair production and inverse Compton scattering governs the spectrum of high-energy radiation in a variety of astrophysical problems [7, 8, 9, 10, 11, 12, 13, 14, 15].

When cascade photons reach energies too low for pair production on the black body photons, the cascade development is slowed, and further development occurs in the gas by ordinary pair production, but with electrons still losing energy mainly by inverse Compton scattering in the black body radiation:

$$\gamma + Z \to Z + e^+ + e^-, \quad e + \gamma_{\rm bb} \to e' + \gamma'. \tag{2}$$

As first pointed out by Lindley [16], since the observed ratios of D/$^4$He and $^3$He/$^4$He are very small ($\sim 10^{-5} - 10^{-4}$), cascade nucleosynthesis can put strong constraints on the energy going into the particles initiating cascades ($\gamma, e^+, e^-$) in the early universe. Cosmological applications of cascade nucleosynthesis have been discussed in several papers [16, 17, 1, 2]. The strongest constraints will be placed for redshifts between $10^3$ and $10^6$. At $z > 3 \times 10^6$ energies of cascade photons are below the threshold for D and $^3$He production, while at $z < 10^3$, direct observation of isotropic X-rays and $\gamma$-rays places more severe limits on the cascade energy deposition.

# 2 The Processes

Electromagnetic cascades in the early universe take place rapidly in the radiation field, and then slowly in the matter. The processes involved in the cascade in the radiation field are photon-photon pair production, inverse Compton scattering and photon-photon scattering. For interactions in the radiation field, the mean interaction rates for all three processes in black body radiation with temperature $T$ are given by (assuming $E \gg mc^2$)

$$\Gamma_{\text{int}}(E, z) = c \int_{\varepsilon_{\min}}^{\infty} n[\varepsilon, T(z)] \int_{-1}^{\mu_{\max}} \frac{\sigma_{\text{int}}(s)(1 - \beta\mu)}{2} d\mu d\varepsilon, \quad (3)$$

where $n(\varepsilon, T)$ is the differential photon number density of the radiation field, $\sigma_{\text{int}}(s)$ is the cross-section for the process, $s$ is the square of the total center of momentum frame energy, $\mu = \cos\theta$ is the cosine of the interaction angle, and $\beta c$ is the velocity of the primary particle ($\beta = 1$ for photons). Here $E$ is the energy of the primary particle and $\varepsilon$ is the target photon energy. For photon-photon scattering and inverse Compton scattering there is no threshold energy, and hence $\varepsilon_{\min} = 0$ and $\mu_{\max} = 1$. The threshold condition for photon-photon pair production is $s > 4m^2c^4$, giving $\varepsilon_{\min} = m^2c^4/E$ and $\mu_{\max} = 1 - 2m^2c^4/(\varepsilon E)$. We assume that the present temperature of the microwave background radiation is $T_0 = 2.735$ K.

Reno and Seckel [18] have explored the consequences of massive particle decay into unstable hadrons during the era of primordial nucleosynthesis. Here, particles in the resulting hadronic cascades interact with nucleons affecting the neutron to proton ratio, and hence changing the relative abundances of $^4$He, $^3$He, D and other light isotopes. For massive particle decay at somewhat later epochs ($\sim 10^3 - 10^7$ s), Dimopoulos et al. [19] have considered the breakup of $^4$He by hadronic cascades. For the epoch under consideration in the present paper ($> 3 \times 10^7$ s), unstable hadrons resulting from decay of massive particles will decay into neutrinos and an electromagnetic component (electrons and photons) before interacting. We therefore only consider the cascade due to the electromagnetic component but include in our later discussion the fraction of the massive particle's rest mass energy carried away by neutrinos.

For interactions in the matter, the following processes are included: ordinary (Bethe-Heitler) pair production on hydrogen and helium, Compton scattering of energetic photons by electrons, photoproduction of pions in photon-proton and photon-helium collisions, and bremsstrahlung by energetic electrons on hydrogen and helium. Since the matter density depends on epoch as $\rho \propto (1+z)^{-3}$, for interactions with matter the interaction rates scale with $z$ as

$$\Gamma_{\text{int}}^{(M)}(E, z) = (1 + z)^3 \Gamma_{\text{int}}^{(M)}(E, z = 0). \quad (4)$$

In the case of radiation, the number density of black body photons, $n_{\text{bb}} = \int n(\varepsilon, T) d\varepsilon$, also depends on epoch as $n_{\text{bb}} \propto (1+z)^{-3}$. However, because the target photon energies also depend on epoch, $\varepsilon \propto (1 + z)$, for interactions with radiation,

$$\Gamma_{\text{int}}^{(R)}(E, z) = (1 + z)^3 \Gamma_{\text{int}}^{(R)}\{(1 + z)E, z = 0\}. \quad (5)$$

Mean interaction rates are illustrated in Fig. 1 for all of the processes discussed above. In the case of interactions with radiation, these are given for the present epoch, $(1+z) = 1$, but they may be scaled to any epoch shifting the corresponding curve by a factor $(1 + z)$ towards lower energies as described in Eq. (5).

# 3 Qualitative Treatment of Electromagnetic Cascades at Large Redshift

The black body radiation at the epochs of interest ($z < 10^7$) is characterized by temperature, $T(z) = 2.735(1 + z)$ K, and photon number density,

$$n_{\rm bb} = 4.22 \times 10^2 T_{2.75}^3 (1 + z)^3 \qquad {\rm cm}^{-3}, \tag{6}$$

where $T_{2.75} = 0.995$ is the temperature in units of 2.75 K.

The density of baryonic gas is

$$\rho_b(z) = 1.88 \times 10^{-29}(1 + z)^3 \Omega_b h^2 \qquad {\rm g\ cm}^{-3} \tag{7}$$

where $\Omega_b = \rho_b/\rho_c$ is the mass fraction of baryons in the universe and $h$ is Hubble's constant in units of 100 km s$^{-1}$ Mpc$^{-1}$. From the recent review by Copi, Schramm and Turner [20], for standard nucleosynthesis one has

$$0.009 \leq \Omega_b h^2 \leq 0.02, \tag{8}$$

and

$$0.4 \leq h \leq 1.0. \tag{9}$$

The baryonic gas consists mainly of hydrogen ($\sim 77\%$ by mass) and helium ($\sim 23\%$ by mass). The radiation length for gas of this composition is $X_0 \approx 66.6$ g cm$^{-2}$.

The characteristic interaction rates for photon-photon pair production (PP) and ordinary pair production, i.e. Bethe-Heitler (BH) process, are

$$\Gamma_{\rm PP}(E, z) \approx 2.2 \times 10^{-12}(1 + z)^3 \sqrt{\frac{E_a(z)}{E}} \exp(-E_a(z)/E) \quad {\rm s}^{-1}, \tag{10}$$

at $E \ll E_a(z)$, where

$$E_a(z) = \frac{m_e c^2}{kT} = \frac{1.12 \times 10^6}{(1 + z)} \quad {\rm GeV}, \tag{11}$$

and, using the asymptotic pair production cross sections,

$$\Gamma_{\rm BH}(E, z) \approx 1.7 \times 10^{-22}(1 + z)^3 \quad {\rm s}^{-1}. \tag{12}$$

We must compare these with the expansion rate

$$H(z) = \begin{cases} 3.24 \times 10^{-18} h(1+z)^{3/2} \quad {\rm s}^{-1} & {\rm for\ } z < 2.3 \times 10^4 \Omega h^2 T_{2.75}^{-4} \\ 1.65 \times 10^{-20} T_{2.75}^{-2}(1+z)^{-2} \quad {\rm s}^{-1} & {\rm for\ } z > 2.3 \times 10^4 \Omega h^2 T_{2.75}^{-4} \end{cases} \tag{13}$$

Strictly, we should use the correct energy-dependent cross sections (described later), but this is sufficient for the present discussion. At high energies, the cascade develops entirely on the black body photons by photon-photon pair production and inverse Compton scattering. The characteristic interaction rates for these processes are much higher than the expansion rate, $H(z)$, and thus one can assume the cascade spectrum is formed instantly. We shall refer to this spectrum as "the zero-generation spectrum".

The zero-generation spectrum extends up to a maximum energy, $E_C$, which is determined at low redshifts ($z < 10^3$) by

$$\Gamma_{\rm PP}(E,z) \approx H(z), \qquad (14)$$

and at high redshifts ($z > 10^3$) by

$$\Gamma_{\rm PP}(E,z) \approx \Gamma_{\rm BH}(E,z). \qquad (15)$$

The maximum energy obtained in this way is given approximately by

$$E_C(z) \approx \begin{cases} 4.5 \times 10^4 (1+z)^{-1} \text{ GeV} & \text{for } z < 10^3 \\ 4.7 \times 10^4 (1+z)^{-1} \text{ GeV} & \text{for } z > 10^3. \end{cases} \qquad (16)$$

For small $z$, the zero-generation spectrum for a cascade of primary energy $E_0$ can be approximately calculated analytically as described by Berezinsky $et$ $al.$ [11] with the result

$$n_\gamma^{(0)}(E) \approx \begin{cases} K_0 (E/E_X)^{-1.5} & \text{for } E < E_X \\ K_0 (E/E_X)^{-2} & \text{for } E_X < E < E_C \\ 0 & \text{for } E_C < E \end{cases} \qquad (17)$$

Here $n_\gamma^{(0)}(E)$ gives the differential photon spectrum, $E_X = 1.78 \times 10^3 (1+z)^{-1}$ GeV is the power-law break-energy appropriate for a black body target photon spectrum, and

$$K_0 \approx \frac{E_0}{E_X^2 [2 + \ln(E_C/E_X)]} \qquad (18)$$

is the normalization constant. This spectrum is confirmed by Monte Carlo simulation at $z = 0$.

When the cut-off energy, $E_C$ given by Eq. (16), is less than the threshold for photodisintegration of $^4$He nuclei ($\sim 20$ MeV), cascade nucleosynthesis is inefficient. This condition restricts the epoch of cascade nucleosynthesis to $z \lesssim 2 \times 10^6$. At large redshifts the spectrum is considerably distorted by $\gamma\gamma \to \gamma\gamma$ scattering, as was first demonstrated by Svensson and Zdziarski [21], and we take this into account in our accurate calculations.

At redshifts between $10^3$ and $2 \times 10^6$, where the subsequent cascade development is via ordinary pair production and inverse Compton scattering, the zero-generation photons survive for a time determined by either the energy loss rate for Compton scattering or the interaction rate for ordinary pair production in the gas. During this time they can produce light nuclei by photodisintegration. The electrons and positrons give rise to first generation photons as a result of inverse Compton scattering. These first generation photons then produce the second generation photons, etc. Each generation of photons is strongly shifted to low energies because the inverse Compton scattering is in the Thomson regime, and only 1 – 2 generations of photons are sufficiently energetic to induce cascade nucleosynthesis.

At redshifts $z \lesssim 10^3$, when interaction times become larger than the Hubble time, only the zero-generation photons are produced, and the effectiveness of cascade nucleosynthesis diminishes as $z$ decreases. At these redshifts, however, cascade photons can be observed directly and such direct observations more strongly constrain the energy density of massive primordial particles.

## 3.1 Approximate calculation of cascade nucleosynthesis

For the redshift range $10^3 - 2 \times 10^6$ the zero generation photons survive ordinary pair production and Compton scattering for a time $[\Gamma_{\rm BH}(E) + k(E)\Gamma_{\rm CS}(E)]^{-1}$ where

$$k(E) \approx 1 - \frac{4/3}{\ln(2E/m_e c^2) + 1/2} \tag{19}$$

is the average fraction of energy lost in Compton scattering. Neglecting subsequent generations, the number of D nuclei produced by the cascade is given approximately by

$$N_D \approx n_{\rm He} c \int [\Gamma_{\rm BH}(E) + k(E)\Gamma_{\rm CS}(E)]^{-1} n_\gamma^{(0)}(E) \sigma_D^{\rm eff}(E) dE \tag{20}$$

where $n_{\rm He}$ is the number density of He nuclei, $\sigma_D$ is the effective cross section for photodisintegration of He into D.

The effective cross section is the sum over partial cross sections of all channels giving rise to the nucleus in question, weighted by the multiplicity. For example, for photodisintegration of $^4$He into D we have

$$\sigma_D^{\rm eff}(E) = \sigma(\gamma, p\,n\,{\rm D}; E) + 2\sigma(\gamma, {\rm D\,D}; E), \tag{21}$$

and for photodisintegration of $^4$He into $^3$He we have

$$\sigma_{^3{\rm He}}^{\rm eff}(E) = \sigma(\gamma, {}^3{\rm He}\,n; E) + \sigma(\gamma, {}^3{\rm H}\,p; E) \tag{22}$$

where we have included production of $^3$H because it decays into $^3$He. For the photodisintegration cross sections, we have used data of Arkatov *et al.* [22]. The effective cross sections used for photodisintegration are plotted in Fig. 2(a). We see that the important photon energy range is between 25 MeV and 100 MeV and that, above threshold, the cross section for production of $^3$He is much higher than for production of D.

In the energy range where photodisintegration of $^4$He is important, the pair production cross sections have not yet reached their asymptotic values, and are strongly energy dependent. We show in Fig. 3 the pair production cross sections for hydrogen and helium for the two cases: fully ionized matter, and neutral matter. For $E < 100$ MeV, the cross sections are almost independent of ionization state, but are quite different from the asymptotic values. The interaction rate for pair production is given by

$$\Gamma_{\rm BH}(E) = [n_H \sigma_{\gamma H}(E) + n_{\rm He}\sigma_{\gamma He}(E)]c. \quad (23)$$

Hence, from Eq. (20) we obtain

$$N_D \approx \int \frac{(Y/4)\sigma_D^{\rm eff}(E)n_\gamma^{(0)}(E)}{(1-Y)\sigma_{\gamma H}(E) + (Y/4)\sigma_{\gamma He}(E) + (1-Y/2)k(E)\sigma_{\rm CS}(E)} dE \quad (24)$$

where $Y$ is the fraction of helium in the early universe by mass, and $\sigma_{\rm CS}(E)$ is the Klein-Nishina cross section. A similar equation governs the number of $^3$He nuclei produced.

# 4 Accurate Calculation

To take account of the exact energy dependences of all the cross sections we use the Monte Carlo method. However, direct application of Monte Carlo techniques to cascades dominated by the physical processes described above over cosmological time intervals presents some difficulties, which we will try to address in the following sections.

The approach we use here is based on the matrix multiplication method described by Protheroe [9] and subsequently developed by Protheroe & Stanev [14]. We use a Monte Carlo program to calculate the yields of secondary particles due to interactions with the thermal radiation and matter. The yields are then used to build up transfer matrices which describe the change in the spectra of particles produced after propagating through the radiation/matter environment for a time $\delta t$. Manipulation of the transfer matrices as described below enables one to calculate the spectra of particles resulting from propagation over arbitrarily large times.

## 4.1 Matrix method

We use 110 fixed logarithmic energy bins of width $\Delta \log E = 0.1$ covering the energy range from $10^{-3}$ GeV to $10^8$ GeV. For example, the energy range of the $j$th energy bin runs from $10^{(j-31)/10}$ GeV to $10^{(j-30)/10}$ GeV. The energy spectra of electrons and photons in the cascade at time $t$ are represented by by vectors $F_j^e(t)$ and $F_j^\gamma(t)$ which give respectively the total number of electrons, and photons, in the $j$th energy bin at time $t$.

The numbers of nuclei produced by photodisintegration of $^4$He nuclei by photons in the cascade are also represented by vectors, $F_j^3(t)$ and $F_j^2(t)$, which give respectively the total number of $^3$He nuclei and D ($^2$H) nuclei produced by interactions of photons having energy in the $j$th energy bin at time $t$. That is, in this case the energy bin index refers to the energy of the photon responsible for the photodisintegration, and not to the energy of the produced nucleus, which is negligible.

We define transfer matrices, $T_{ij}^{\mu\nu}(\delta t)$, which give the number of particles of type $\nu = e$ (electron), $\gamma$ (photon), 3 ($^3$He) or 2 (deuterium) in the bin $j$ which result at a time $\delta t$ after a particle of type $\mu =$ e or $\gamma$ and energy in the bin $i$ initiates a cascade. Then, given the spectra of particles at time $t$ we can obtain the spectra at time $(t + \delta t)$

$$F_j^{\rm e}(t + \delta t) = \sum_{i=j}^{110} \left[ {\rm T}_{ij}^{\rm ee}(\delta t) F_i^{\rm e}(t) + {\rm T}_{ij}^{\gamma \rm e}(\delta t) F_i^{\gamma}(t) \right], \quad (25)$$

$$F_j^{\gamma}(t + \delta t) = \sum_{i=j}^{110} \left[ {\rm T}_{ij}^{\rm e\gamma}(\delta t) F_i^{\rm e}(t) + {\rm T}_{ij}^{\gamma\gamma}(\delta t) F_i^{\gamma}(t) \right], \quad (26)$$

$$F_j^3(t + \delta t) = \sum_{i=j}^{110} \left[ {\rm T}_{ij}^{\rm e3}(\delta t) F_i^{\rm e}(t) + {\rm T}_{ij}^{\gamma 3}(\delta t) F_i^{\gamma}(t) \right], \quad (27)$$

$$F_j^2(t + \delta t) = \sum_{i=j}^{110} \left[ {\rm T}_{ij}^{\rm e2}(\delta t) F_i^{\rm e}(t) + {\rm T}_{ij}^{\gamma 2}(\delta t) F_i^{\gamma}(t) \right], \quad (28)$$

where $F_i^{\rm e}(t)$ and $F_i^{\gamma}(t)$ are the input electron and photon spectra (number of electrons or photons in the $i$th energy bin).

We could also write this as

$$[{\rm F}(t + \delta t)] = [{\rm T}(\delta t)][{\rm F}(t)] \quad (29)$$

where

$$[{\rm F}] = \begin{bmatrix} F^{\rm e} \\ F^{\gamma} \\ F^3 \\ F^2 \end{bmatrix}, \quad [{\rm T}] = \begin{bmatrix} {\rm T}^{\rm ee} & {\rm T}^{\gamma \rm e} & 0 & 0 \\ {\rm T}^{\rm e\gamma} & {\rm T}^{\gamma\gamma} & 0 & 0 \\ {\rm T}^{\rm e3} & {\rm T}^{\gamma 3} & {\rm I} & 0 \\ {\rm T}^{\rm e2} & {\rm T}^{\gamma 2} & 0 & {\rm I} \end{bmatrix}. \quad (30)$$

## 4.2 Transfer matrix calculation

The transfer matrices depend on particle yields, $Y_{ij}^{\alpha\beta}$, which we define as the probability of producing a particle of type $\beta$ in the energy bin $j$ when a primary particle with energy in bin $i$ undergoes an interaction of type $\alpha$. To calculate $Y_{ij}^{\alpha\beta}$ we use a Monte Carlo simulation. For inverse Compton scattering and photon-photon pair production we have used the computer code described by Protheroe [9, 12], updated to model interactions with a thermal photon distribution of arbitrary temperature. For photon-photon scattering, we have used the cross sections given by Berestetskii et al. [23].

In the case of inverse Compton scattering in the Thomson regime, the basic matrix method fails to predict correctly the electron spectrum, and hence the emitted photon spectrum. This is because the fraction of energy lost per interction is small, and effectively the electrons suffer continuous energy losses,

$$\frac{dE}{dt} = -bE^2. \quad (31)$$

Injection of one electron with energy $E_0$ at $t = 0$ should result in one electron with energy

$$E(t) \approx (bt + E_0^{-1})^{-1} \quad (32)$$

at time $t$ (i.e. there is very little spread in the final energy). However, the basic matrix method would give rise to a broad energy distribution with mean energy equal to $E(t)$. In the present problem, electrons in the Thomson regime lose energy by inverse Compton scattering at a rate very much higher than by the competing process, bremsstrahlung. Therefore we immediatly replace any electron produced with energy in the Thomson regime, or in the transition region between Thomson and Klein-Nishina regimes, by all the photons from inverse Compton scattering that would be produced while the electron subsequently cools. We define a matrix $G_{ij}^{\text{IC}}$ which gives the number of photons produced in energy bin $j$ by an electron injected with energy in bin $i \leq m$. We have chosen the maximum energy of electrons treated in this way to be just below the photon-photon pair production threshold, i.e. $m = 77$ at the present epoch. For $61 \leq i \leq 77$ (at the present epoch) the Monte Carlo method is used to calculate $G_{ij}^{\text{IC}}$ for $j \geq 61$, and the remaining array elements, for which the electron is well inside the Thomson regime, are obtained from the distribution function which we obtain by numerical integration,

$$\frac{dN}{dE} = \frac{9m_e c^2 \int_0^\gamma d\gamma \gamma^{-4} \int d\varepsilon \varepsilon^{-1} n(\varepsilon) f(\hat{\varepsilon}_1)}{16 \int d\varepsilon \varepsilon n(\varepsilon)}, \tag{33}$$

where $f(\hat{\varepsilon}_1)$ is given in ref. [24].

For interactions with matter (ordinary pair production, bremsstrahlung, Compton scattering, photoproduction) we assume a composition 77% hydrogen and 23% helium by mass. Photon and electron yields from photoproduction follow from the $\pi^\pm \to \mu^\pm \to e^\pm$ and $\pi^0 \to 2\gamma$ decays. In the case of ordinary pair production and bremsstrahlung, yields are calculated for both neutral and fully ionized matter, and mixed in a ratio appropriate to the fractional ionization at the epoch for which the transfer matrices are to be calculated. For the fractional ionization we use Eq. (3.95) of Kolb and Turner [25].

It is important to emphasize that in the most important energy range (20 – 100 MeV) the cross–sections for bremsstrahlung and pair production are very strongly energy dependent. For pair production we use the direct calculations of Hubbel, Gimm and Overbo [26] for hydrogen and helium, in which the screening correction (for neutral matter) is explicit. For bremsstrahlung we use the expressions of Koch and Motz [27] with form factors for hydrogen and helium adjusted to represent the more precise values of Tsai [28]. The cross-sections for fully ionized matter are calculated with the bremsstrahlung formulae valid in the absence of screening by the atomic electrons.

In the case of production of $^3$He and deuterium nuclei, the yield is simply the effective cross section for production of the nucleus in question divided by the total cross section. There is, however, a further complication. During pion photoproduction on $^4$He the nucleus almost always fragments, and we must therefore take account of photodisintegration during pion photoproduction. We know of no experimental data giving branching ratios for the various possible final state nuclei, and new measurements are urgently required. In the mean time, we assume this process to be similar to the break-up of $^4$He during collisions with nucleons in which pions are produced. Following Meyer [29] we assume final states ($^3$He $n$) : ($^3$H $p$) : (D D) : ($p\,n$ D) are produced in the ratio 2:2:1:2 with neg-ligible production of the final state $(2p\,2n)$. Hence the effective cross sections used for photodisintegration during pion photoproduction are

$$\sigma_{\text{eff}}(\text{D, pions}) = \sigma_{\text{eff}}(^3\text{He, pions}) = \frac{4}{7}\sigma(\text{photoproduction}). \quad (34)$$

The effective cross sections used for photodisintegration during pion photoproduction have been added to Fig. 2(a). In Fig. 2(b) we have plotted the effective cross sections for both photodisintegration and disintegration during photoproduction multiplied by $E^{-0.7}$. For this log–linear plot, the areas then show the relative contributions to $^3$He and D production for an $E^{-1.7}$ photon spectrum (a single power-law approximation to the photon spectrum in the cascade). Clearly, disintegration during photoproduction is particularly important for calculating abundances of deuterium, but is less important for $^3$He production. Since the inferred ratio of ($^3$He + D) to $^4$He is of the order of $10^{-4}$, this implies that photodisintegration of $^3$He and D is unimportant in determining their abundances. We therefore neglect this process.

To calculate the transfer matrices we have used a modification of the semi-analytical technique described by Protheroe & Stanev [14]. From Fig. 1 we see that at all epochs the highest interaction rate is that for inverse Compton scattering by electrons at low energies where the scattering is in the Thomson regime, $\Gamma_{\text{IC}} \to \Gamma_{\text{T}} = n_{\text{bb}}\sigma_T c$. If $\delta t$ is much shorter than the shortest interaction time in the cascade, i.e. $\delta t \ll 1/\Gamma_{\text{T}}$, then

$$T_{ij}^{\text{ee}}(\delta t) \approx \delta_{ij}[1 - \delta t \Gamma_e(E_i)] + \delta t[\Gamma_{\text{IC}}(E_i) Y_{ij}^{\text{IC e}} + \Gamma_{\text{brem}}(E_i) Y_{ij}^{\text{brem e}}], \quad (35)$$

$$T_{ij}^{\text{e}\gamma}(\delta t) \approx \delta t[\Gamma_{\text{IC}}(E_i) Y_{ij}^{\text{IC }\gamma} + \Gamma_{\text{brem}}(E_i) Y_{ij}^{\text{brem }\gamma}], \quad (36)$$

$$T_{ij}^{\gamma\text{e}}(\delta t) \approx \delta t[\Gamma_{\text{PP}}(E_i) Y_{ij}^{\text{PP e}} + \Gamma_{\text{CS}}(E_i) Y_{ij}^{\text{CS e}} + \Gamma_{\text{BH}}(E_i) Y_{ij}^{\text{BH e}} + \Gamma_{\text{photo}}(E_i) Y_{ij}^{\text{photo e}}], \quad (37)$$

$$T_{ij}^{\gamma\gamma}(\delta t) \approx \delta_{ij}[1 - \delta t \Gamma_\gamma(E_i)] +$$
$$+ \delta t[\Gamma_{\text{scat}}(E_i) Y_{ij}^{\text{scat }\gamma} + \Gamma_{\text{CS}}(E_i) Y_{ij}^{\text{CS }\gamma} + \Gamma_{\text{photo}}(E_i) Y_{ij}^{\text{photo }\gamma}], \quad (38)$$

$$T_{ij}^{\text{e}3}(\delta t) \approx 0, \quad (39)$$

$$T_{ij}^{\text{e}2}(\delta t) \approx 0,, \quad (40)$$

$$T_{ij}^{\gamma 3}(\delta t) \approx \delta t[\Gamma_{\text{PD}}(E_i) Y_{ij}^{\text{PD 3}} + \Gamma_{\text{photo}}^{(\text{He})}(E_i) Y_{ij}^{\text{photo 3}}], \quad (41)$$

$$T_{ij}^{\gamma 2}(\delta t) \approx \delta t[\Gamma_{\text{PD}}(E_i) Y_{ij}^{\text{PD 2}} + \Gamma_{\text{photo}}^{(\text{He})}(E_i) Y_{ij}^{\text{photo 2}}], \quad (42)$$

$$T_{ij}^{33}(\delta t) = \delta_{ij}, \quad (43)$$

$$T_{ij}^{22}(\delta t) = \delta_{ij} \quad (44)$$

where

$$\Gamma_e(E_i) = \Gamma_{\text{IC}}(E_i) + \Gamma_{\text{brem}}(E_i), \quad (45)$$

and

$$\Gamma_\gamma(E_i) = \Gamma_{\text{PP}}(E_i) + \Gamma_{\text{scat}}(E_i) + \Gamma_{\text{CS}}(E_i) + \Gamma_{\text{photo}}(E_i) + \Gamma_{\text{BH}}(E_i) + \Gamma_{\text{PD}}(E_i) \quad (46)$$

give the total interaction rates of electrons and photons. In the equations above, we have used the following abbreviations: IC (inverse Compton), brem (bremsstrahlung),

PP (photon-photon pair production), CS (Compton scattering), BH (Bethe-Heitler pair production), scat (photon-photon scattering), photo (photoproduction on both hydrogen and helium), photo (He) (photoproduction on helium), and PD (photodisintegration).

Modifications outlined earlier to take account of inverse Compton scattering properly in the Thomson regime and the transition region between the Thomson and Klein-Nishina regimes (i.e. for $j \leq m$) are as follows:

$$T_{ij}^{e\gamma}(\delta t) \leftarrow T_{ij}^{e\gamma}(\delta t) + \sum_{k=1}^{m} \delta t [\Gamma_{\text{IC}}(E_i) Y_{ik}^{\text{IC e}} + \Gamma_{\text{brem}}(E_i) Y_{ik}^{\text{brem e}}] G_{kj}^{\text{IC}}, \qquad (47)$$

$$T_{ij}^{\gamma\gamma}(\delta t) \leftarrow T_{ij}^{\gamma\gamma}(\delta t) + \sum_{k=1}^{m} \delta t [\Gamma_{\text{PP}}(E_i) Y_{ik}^{\text{PP e}} + \Gamma_{\text{CS}}(E_i) Y_{ik}^{\text{CS e}} +$$

$$+ \Gamma_{\text{BH}}(E_i) Y_{ik}^{\text{BH e}} + \Gamma_{\text{photo}}(E_i) Y_{ik}^{\text{photo e}}] G_{kj}^{\text{IC}} \qquad (48)$$

$$T_{ij}^{ee}(\delta t) = \delta_{ij}[1 - \delta t \Gamma_e(E_i)] \qquad (49)$$

$$T_{ij}^{\gamma e}(\delta t) = 0. \qquad (50)$$

We require $1/\delta t$ be much larger than the largest interaction rate in the problem, $\Gamma_T \approx 10^{-11}(1+z)$ s$^{-1}$, and hence typically we use $\delta t \approx 10^{10}/(1+z)$ s. The cascade is followed for a time $t_{\max}$ which must be much longer than the largest interaction time for interactions with matter at the energies at which photodisintegration and disintegration during photoproduction can be significant, say from 30 MeV to 10 GeV. At about 10 GeV, $\Gamma_{\text{CS}} \approx 2 \times 10^{-25}(1+z)$ s$^{-1}$ so we would require $t_{\max} \sim 5 \times 10^{25}/(1+z)$ s. To complete the calculation of the cascade over time $t_{\max}$ using repeated application of the transfer matrices would therefore require $t_{\max}/\delta t \sim 5 \times 10^{15}$ steps. This is clearly impractical, and we must use the more sophisticated approach described below.

## 4.3 Matrix doubling method

The matrix method and matrix doubling technique have been used for many years in radiative transfer problems [30, 31]. The method used here to calculate the spectrum of particles emerging after an arbitrary time is that described by Protheroe & Stanev [14], and is summarized below. Once the transfer matrices have been calculated for a time $\delta t$, the transfer matrix for a time $2\delta t$ is simply given by applying the transfer matrices twice, i.e.

$$[T(2\delta t)] = [T(\delta t)]^2. \qquad (51)$$

In practice, it is necessary to use double precision and to ensure that energy conservation is preserved after each doubling. The new matrices may then be used to calculate the transfer matrices for a time interval $4\delta t$, and so on. A time interval $2^n \delta t$ only requires the application of this 'matrix doubling' $n$ times. The spectrum of electrons and photons after a large time interval $\Delta t$ is then given by

$$[F(t + \Delta t)] = [T(\Delta t)][F(t)] \qquad (52)$$

where [F(t)] represents the input spectra, and $\Delta t = 2^n \delta t$. In this way, cascades over long time intervals can be modelled quickly and efficiently. For example, to simulate the cascade over $t_{\max} \sim 5 \times 10^{25}/(1+z)$ s with an initial step size of $\delta t \sim 10^{10}/(1+z)$ s would take only 52 steps.

As a test we have run the program over such large time intervals and switched off all processes except photon-photon pair production, inverse Compton scattering and photon-photon scattering so that our results could be compared directly with those of Svensson and Zdziarski [21]. In their calculations, Svensson and Zdziarski continuously inject into a radiation field photons or electrons with energies above the threshold for photon-photon pair production, and solve the kinetic equation to find the steady-state spectrum, for some given constant escape time $t_{\rm esc}$, which they refer to as the "escaping spectrum". Two important energies enter in the problem, the maximum photon energy $E_m$ defined as the energy at which the photon-photon pair production rate equals the escape rate, and energy $E_c$ defined as the energy at which the photon-photon scattering rate equals the escape rate:

$$\Gamma_{\rm PP}(E_m) = t_{\rm esc}^{-1}, \qquad (53)$$

$$\Gamma_{\rm scat}(E_c) = t_{\rm esc}^{-1}. \qquad (54)$$

For the case of a black body radaition field Svensson and Zdziarski show results of their calculations, both with and without photon-photon scattering, for the case where $E_m/E_c = 4$. In order for us to compare with these results, it was necessary first to work out the value of $t_{\rm esc}$ such that $E_m/E_c = 4$ for black body radiation of a given temerature. The matrix program was then run for injection of a primary electron of energy $E_i \gg E_m$ to obtain the spectrum of photons time $t$ after injection, $F^\gamma(E,t)$, for $0 < t \leq 10 t_{\rm esc}$. We then obtain the escaping spectrum per energy injected by integration,

$$F_{\rm esc}^\gamma(E) = \frac{1}{E_i t_{\rm esc}} \int_0^{t_{\max}} F^\gamma(E,t) \exp\left(-\frac{t}{t_{\rm esc}}\right) dt. \qquad (55)$$

In Fig. 4 we plot $\eta(E) = E^2 F_{\rm esc}^\gamma(E)$ against $E/E_m$ for our calculations both including and neglecting photon-photon scattering. Our results are compared with those of Svensson and Zdziarski [21]. Note that the results of Svensson and Zdziarski are arbitrarily normalized such that the no-scattering case has $\eta(E_m) = 1$. We see that the agreement is satisfactory. We also show the spectrum (Eq. 17) used in our approximate treatment and the spectrum given in Equation 11 of Ellis *et al.* [1].

## 5 Redshifting

For $(1+z) \gtrsim 10^5$, we must take account of redshifting because the expansion rate of the universe, $H$, becomes comparable to the interaction rates in matter. The approach we adopt is to propagate initially over a time interval $\Delta t$ which would give rise to a change in $\log(1+z)$ equal to the width of the energy bins,

$$\Delta \log(1+z) = 0.1. \qquad (56)$$

We then redshift the energy bin contents of the vectors representing the photon and electron spectra:

$$F_i^\gamma \leftarrow F_{i+1}^\gamma; \quad F_i^e \leftarrow F_{i+1}^e. \tag{57}$$

Much of the cascade development takes place in this first interval. A further application of the transfer matrices for further propagation over an additional time $\Delta t$ would give rise to a change in $\log(1+z)$ approximately equal to the width of the energy bins. We then redshift the energy bin contents of the vectors representing the electron and photon spectra as described above. While not strictly exact, because the further redshift change does not correspond exactly to $\Delta \log(1+z) = 0.1$, we have found that the error induced by this procedure is insignificant in the present problem. This procedure is repeated until the cascade is complete, and because the redshifting means that particle energies are more rapidly reduced below interaction thresholds, the cascade finishes earlier than without redshifting.

# 6 Results and Discussion

We have performed the cascade calculation to find the number of deuterium and $^3$He nuclei produced by a cascade initiated at the epoch of redshift $z_c$. We give the number of nuclei per 1 GeV of total cascade energy, so that the total number of nuclei is obtained by multiplying by the total energy of the cascade in GeV. Because of the almost instant formation of the zero-generation spectrum, the exact shape of the $\gamma$-ray or electron injection spectrum is of no consequence for further cascade development, and only the total amount of injected energy is relevant. The results are given for redshifts at which cascades are initiated, $z_c$, in the range $10^2 - 10^7$, and for the following values of $h$ (and $\Omega_b$): 0.4 (0.125); 0.7 (0.025); and 1.0 (0.01). The resulting number of $^3$He and D produced per GeV of the total cascade energy is given in Fig. 5(a) and Fig. 5(b) respectively. At redshifts less than $\sim 10^5$ for which the maximum energy of photons in the cascade exceeds the thresholds for production of $^3$He and D by photodisintegration of $^4$He, approximately ten times as many $^3$He nuclei are produced compared with D nuclei. This is to be expected given the cross sections for photodisintegration and disintegration during photoproduction (see Fig. 2).

In Fig. 6 we show the sensitivity of the results to: assumptions about the effective cross sections for photodisintegration; use (or neglect) of the energy energy dependence of the pair production and bremsstrahlung cross sections; and inclusion (or neglect) of photon-photon scattering. In this comparison, we use the $h = 0.7$ results. The full curve is the accurate calculation including all effects. The dotted curve shows the effect of neglecting photon-photon scattering, which is seen to have the greatest effect on the results for $z_c > 3 \times 10^4$. The dashed curve shows the effect of neglecting the disintegration of $^4$He during pion photoproduction and is seen to be negligible for $^3$He production, but accounts for up to 50% of all deuterium production, depending on $z_c$. The dot-dash curve shows the effect of neglecting the disintegration of $^4$He during pion photoproduction, using asymptotic pair production and bremsstrahlung cross sections, and neglecting photon-photon scattering. Finally, The heavy dot-dot-dot-dash curve gives the result of the approximate treatment. Here, we simply apply Eq. (24) using the energy-dependent cross sections, include disintegration during photoproduction, and use the photon spectrum given by Eq. (17).

We are now in a position to summarize the cascade nucleosynthesis scenario in the context of the results given in Figs. 5 and 6. The epoch of cascade nucleosynthesis is limited by $z_{\max}$ and $z_{\min}$. The maximum redshift is determined by the condition that the maximum energy of photons in the cascade spectrum, given by Eq. (17), must be larger than the threshold for D or $^3$He production on $^4$He, $E_{\text{th}} \approx 20\ MeV$. This condition results in

$$z_{\max} \approx 2.4 \times 10^6, \tag{58}$$

which is clearly observed in Figs. 5 and 6.

These Figures demonstrate also that the effectiveness of D and $^3$He production decreases as $z$ decreases. The reason is not the decrease in the density of $^4$He, as one naively might think, but rather the decrease in the number of *low-energy* photons in the cascade. In fact, the lower the redshift, the higher the photon energies in the cascade (both $E_X$ and $E_C$ increase), and therefore a smaller fraction of the photons participates in cascade nucleosynthesis. One can understand this in a semi-qualitative way from Eq. (20). If we neglect the energy dependence of $\Gamma_{\text{BH}}$, and use $n_\gamma^{(0)}(E)$ given by Eq. (17), we obtain

$$\frac{N_D}{E_0} = n_{\text{He}}(z) \frac{x_{\text{rad}}}{\rho_b(z)} \frac{E_X^{-0.5}(z)}{[2 + \ln(E_C/E_X)]} \int_{E_{\text{th}}}^{\infty} E^{-1.5} \sigma_D(E) dE. \tag{59}$$

In deriving this Equation, we took into account that, at small $z$, only the low-energy part of the spectrum (Eq. 17) effectively takes part in nucleosynthesis ($E_X$ is high). From Eq. (59) we see that the factor $(1+z)^3$ in $n_{\text{He}}(z)$ is compensated for by the same factor in $\rho_b(z)$, and then we are left with a factor $(1+z)^{-0.5}$ coming from $E_X(z)$. Numerically, the $(1+z)^{-0.5}$ dependence for the curves in Figs. 5 and 6 would give rather a bad fit because it neglects the dependence of $\Gamma_{\text{BH}}$ on $E$ and the cascading in the gas.

At $z < 10^3$ the cascade photons can be directly observed, and the upper limit for the isotropic gamma-ray flux at $10 - 200$ MeV is more restrictive for the cascade production than nucleosynthesis. Therefore, the most effective epoch for cascade nucleosynthesis corresponds to redshifts in the range $10^3 - 2 \times 10^6$.

From Fig. 6 one can see that that role of $\gamma\gamma \to \gamma\gamma$ scattering is important only for epochs with redshifts $z > 5 \times 10^4$. This is easy to understand from Fig. 1 which shows that, at $z = 0$, $\Gamma_{\text{scat}}$ is practically everywhere below the curves $\gamma p \to$ ee (ordinary pair production) and $\gamma\gamma \to$ ee (photon-photon pair production). For the epoch with redshift $z$ we must coherently shift both curves ($\gamma\gamma \to \gamma\gamma$ and $\gamma\gamma \to$ ee) by a factor $(1+z)$ to the left. At redshifts $z \geq 10^5$ the crossing point of the $\gamma\gamma \to \gamma\gamma$ and $\gamma p \to$ ee curves is at $E <$ 1 GeV where the $\gamma p \to$ ee interaction rate is less because of the energy dependence of the cross section near threshold, and so $\gamma\gamma$ scattering becomes relatively more important. Passing through the energies where $\gamma\gamma \to \gamma\gamma$ scattering dominates, the spectrum of the cascade changes, the main effect being that the cut-off energy in the spectrum is lowered. The scattered photons do not interact again with the target photons; they are just redistributed over the spectrum producing a small bump before the cut off. Both these effects (bump and early cut-off) are prominent only at high redshifts $z \geq 10^5$ for the reasons given above.

We distinguish the zero-generation cascade from the cascades of the first, second, etc., generations. The zero-generation cascade develops on the black body photons mainly due to $\gamma + \gamma_{\rm bb} \to e^+ + e^-$ and $e + \gamma_{\rm bb} \to e + \gamma$ scattering. The characteristic times for these processes are much shorter than for all other processes, and one can assume that the zero-generation spectrum is formed instantly. At small $z$ the spectrum is given by Eq. (17); at large $z$ it is distorted by $\gamma\gamma \to \gamma\gamma$ scattering, and by energy dependence of the inverse Compton scattering.

At large redshift, $z \sim z_{\rm max}$, only photons from the zero-generation cascade participate in cascade nucleosynthesis. In this case, the maximum cascade energy, $E_C(z)$, is close to the threshold of the nuclear reactions. The photons of the first generation are strongly shifted towards low energies, and thus they become sterile. From Fig. 6 one can see that the approximate calculation (dot-dot-dot-dash curve), with the zero-generation photons only, is extremely close to the exact calculation in which $\gamma\gamma \to \gamma\gamma$ scattering is neglected (dotted curve).

The cascade in the gas develops due to $\gamma + Z \to Z + e^+ + e^-$ and $e + \gamma_{\rm bb} \to e + \gamma$ scattering. In the latter process, the scattered photon is strongly shifted to lower energies in comparison with the initial electron because the inverse Compton scattering is in the Thomson regime. As a result, at $z \sim 10^3$ only two generations of photons are sufficiently energetic to induce cascade nucleosynthesis. On the other hand, for cascading in the gas we require

$$\int_t^{t_0} \rho_b c\, dt \geq X \tag{60}$$

where $X = \frac{9}{7}X_0$ is the interaction length in g cm$^{-2}$ ($X_0$ is the radiation length). This implies

$$z \geq \left(\frac{3}{2}\frac{H_0}{c}\frac{X}{\rho_c \Omega_b}\right)^{2/3} = 1.2 \times 10^3 \left(\frac{0.01}{\Omega_b h^2}\frac{h}{0.75}\right)^{2/3}. \tag{61}$$

Therefore for $z \ll 10^3$, again only the zero-generation photons take part in nucleosynthesis.

## 7  Applications and Conclusions

A few words about some applications of our results are now in order. Cascade nucleosynthesis strongly restricts high energy processes at $10^3 < z < 10^6$. From Fig. 5, the number of $^3$He and D nuclei produced by each cascade with total energy $E_0$ (in GeV) is $N(^3{\rm He}) \sim (0.1-1)E_0$ and $N({\rm D}) \sim (1-6) \times 10^{-3} E_0$, where $E_0$ is in GeV. The total production of $^3$He and D during the cascade nucleosynthesis must be less than the observed primordial quantities.

The physical processes of interest include the decay of heavy relic particles [1, 2], the cusp radiation of cosmic strings [3], and the massive particle production by superconducting cosmic strings [4, 5, 6].

We shall give here two examples of applications. In the first example, we obtain the limit on the density of long-lived particles $X$ which can decay into a cascade-producing particle $c$ ($X \rightarrow c+$ anything) with a branching ratio $b_c$. It is easy to calculate the fraction ($^3$He + D)/H at $z = 0$ as

$$\frac{^3\text{He} + \text{D}}{\text{H}} = f_c b_c \frac{n_X^*}{n_H} \left(\frac{m_X c^2}{1\,\text{GeV}}\right) [N(^3\text{He}, \tau_X) + N(\text{D}, \tau_X)] \qquad (62)$$

where $n_X^*$ is the space density that $X$ particles would have at $z = 0$ if they were stable, all densities $n_H$ and $n_X^*$ are taken at $z = 0$, $f_c$ is the fraction of mass $m_X$ transferred to cascade energy, $\tau_X$ is the lifetime of the $X$ particle and

$$N(^3\text{He}, \tau_X) = \frac{1}{\tau_X} \int N(^3\text{He}, z_c(t)) \exp(-t/\tau_X) dt. \qquad (63)$$

Here $N(^3\text{He}, z_c)$ is the number of $^3$He nuclei produced per GeV of the total cascade energy, given in Fig. 5(a). $N(^3\text{He}, \tau_X)$ obtained by convolving the data of Figure 5 with an exponential distribution of decay times with mean decay time $\tau_X$ (Eq. 63) is given in Fig. 7(a). $N(\text{D}, \tau_X)$, given in Fig. 7(b), is obtained from $N(\text{D}, t(z_c))$, given in Fig. 5(b), in exactly the same way.

Defining

$$\Omega_X^* = \frac{n_X^* m_X}{\rho_C}, \qquad (64)$$

assuming baryonic matter is 77% hydrogen by mass, and rearranging Eq. (62) we obtain

$$\Omega_X^*(\tau_X) = \left(\frac{0.77\Omega_b}{f_c b_c}\right) \left(\frac{1\,\text{GeV}}{m_H c^2}\right) \left(\frac{^3\text{He} + \text{D}}{\text{H}}\right) [N(^3\text{He}, \tau_X) + N(\text{D}, \tau_X)]^{-1}. \qquad (65)$$

In Fig. 8 we use the results of Fig. 7 for the total $^3$He plus D produced per GeV of cascade energy, together with the upper limit inferred from measurements of $^3$He in meteorites and the solar wind making assumptions about stellar processing and galactic chemical evolution[20] ($^3$He + D)/H $< 1.1 \times 10^{-4}$, to obtain an upper limit to $\Omega_X^*(\tau_X)$. Note, however, that very recent measurements of D/H are closer to $2.5 \times 10^{-4}$ [32, 33], and if such higher values for ($^3$He + D)/H were adopted, the upper limits we would derive would be correspondingly higher. Since $\Omega_X^*(\tau_X)$ is proportional to $\Omega_b/f_c b_c$, we have plotted $f_c b_c \Omega_X^*(\tau_X)/\Omega_b$. In Fig. 8 we also plot the result of Ellis et al. [1] and a result we would have obtained if we had uses the asymptotic Bethe-Heitler pair production and bremsstrahlung cross sections. The discrepancy between our results and those of Ellis et al. can be partly explained in terms of the cross section. Note that $\Omega_X^*(\tau_X)$ depends on ($^3$He + D), i.e. mainly on $^3$He which is insensitive to assumptions about $^4$He disintegration during pion photoproduction. From Fig. 8 we find that for mean decay times $\tau_X$ ranging from 1 year to $10^6$ years, a density of dark matter in the form of massive primordial particles of only 0.1% to 0.3% of that of normal matter could account for all of the observed $^3$He and deuterium.

As a second example, we obtain an upper limit for the neutrino intensity produced by the decay of $X \to \nu +$ anything with a branching ratio $b_\nu$. We shall assume, as in the first example, that these particles decay also into cascade-producing particles ($X \to c +$ anything) with a branching ratio $b_c$. Since each $X$ particle decay results in $b_\nu$ neutrinos, the present density ($z = 0$) of neutrinos is $n_\nu = b_\nu n_X^*$. Putting $n_X^* = b_\nu^{-1} n_\nu$ into Eq. (62) and using $(^3\text{He} + \text{D})/\text{H} < 1.1 \times 10^{-4}$ one obtains an upper limit for the integral neutrino intensity $I_\nu = (c/4\pi) n_\nu$,

$$I_\nu < 2.3 \Omega_b h^2 \frac{b_\nu}{f_c b_c} \left(\frac{1 \text{ GeV}}{m_X c^2}\right) [N(^3\text{He}, \tau_X) + N(\text{D}, \tau_X)]^{-1}. \quad \text{cm}^{-2} \text{ s}^{-1} \text{ sr}^{-1} \qquad (66)$$

where $N(^3\text{He}, \tau_X)$ and $N(\text{D}, \tau_X)$ are given in Figs. 7(a) and 7(b).

Having only one free parameter (baryonic density), the standard big bang nucleosynthesis is in beautiful agreement with the observations of $^4$He, $^3$He, D and Li, as well as with 3 neutrino flavors observed at LEP. Hence, our cascade nucleosynthesis calculations give upper limits to any hypothetical high energy process at $10^3 < z < 10^6$. However, one should not forget about the potential for cascade nucleosynthesis to give rise to small corrections for standard big bang nucleosynthesis in the production of $^3$He and D. In order to assist such small corrections, we show in Fig. 9 the ratio of D to $^3$He production by cascade nucleosynthesis as a function of both $z_c$ and $\tau_X$.

# 8 Acknowledgments

We are grateful to Subir Sarkar and an anonymous referee for helpful comments on the original manuscript. The research of RJP is supported by a grant from the Australian Research Council. The research of TS is supported in part by DOE Grant DE-FG-91ER40626.

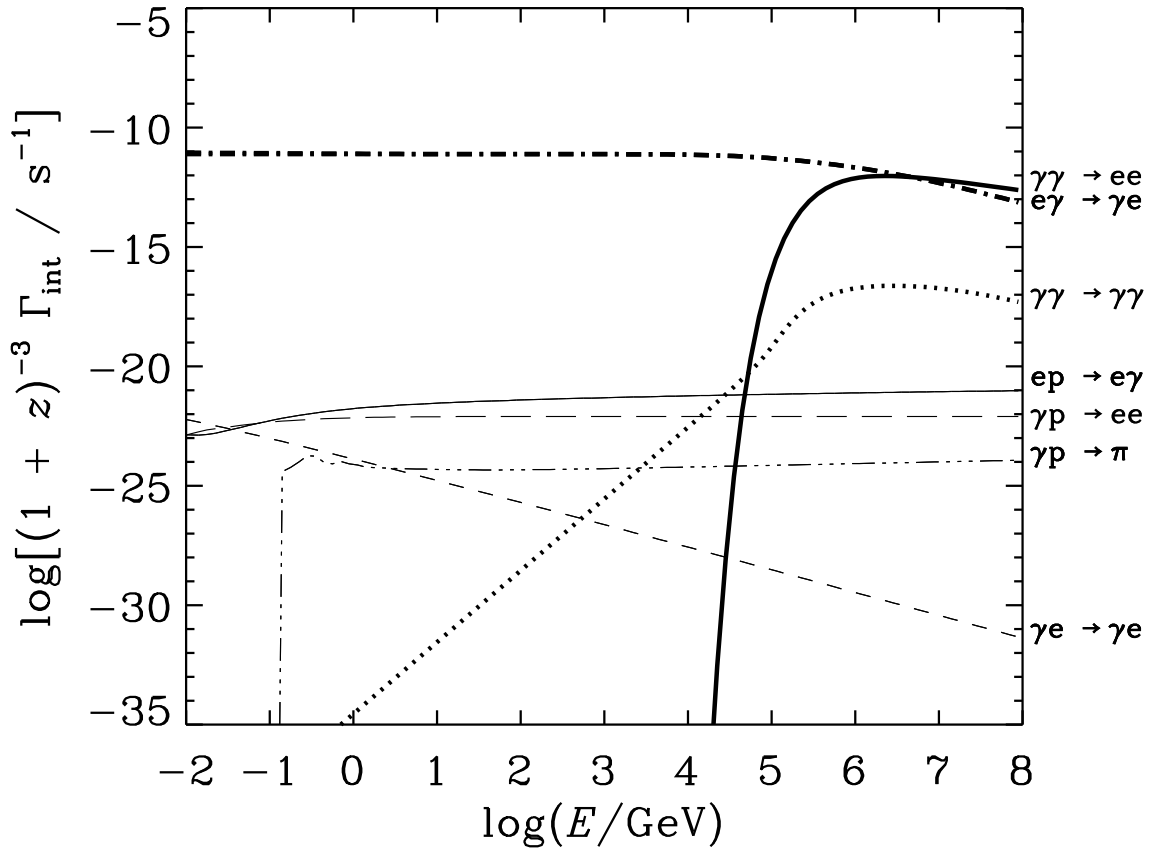

Figure 1: Mean interaction rates at for all of the cascade processes considered in the present work. Interactions with radiation at $z = 0$ are shown by heavy lines: photon-photon pair production (full line); inverse Compton scattering (dot-dash line); photon-photon scattering (dotted line). For other epochs, the interaction rates for particles of energy $E$ are obtained by reading the rates at energy $(1+z)E$, i.e. by shifting the corresponding curve by a factor $(1+z)$ to lower energies. Interactions with fully ionized matter are shown by the thin lines: ordinary pair production (long-dashed line); bremsstrahlung (full line); Compton scattering (short-dashed line); pion photoproduction (dot-dot-dot-dash line). Note: we assume $h = 0.7$ and $\Omega_b = 0.025$ for interactions in matter; for this plot we use fully ionized matter (even though matter is neutral at $z = 0$) because, at most redshifts we consider in this paper, the matter is fully ionized.

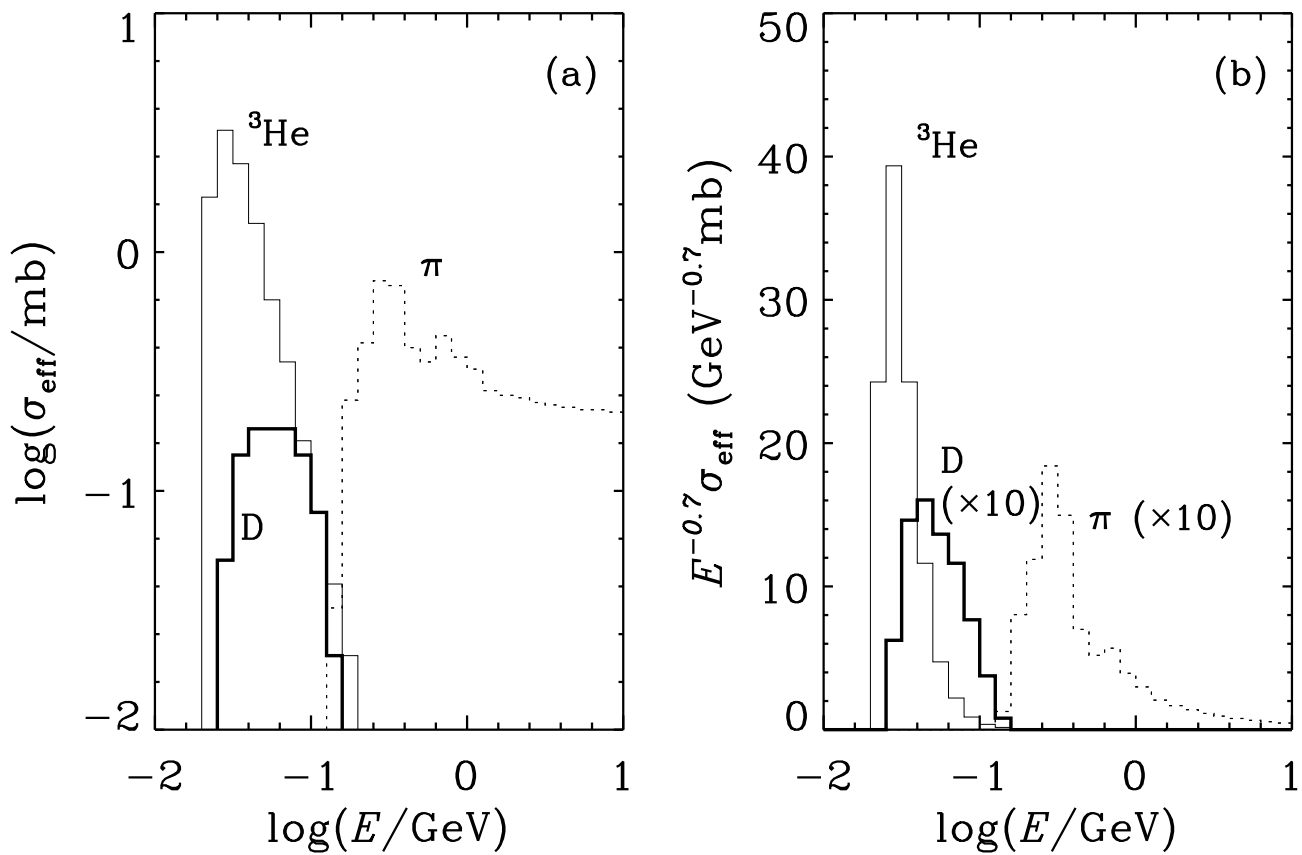

Figure 2: (a) The effective cross sections used for photodisintegration of $^4$He into $^3$He (thin histogram) and D (thick histogram). We estimate that the effective cross sections for disintegration of $^4$He during photoproduction into $^3$He and into D are approximately equal; the effective cross section for either process is shown by the dotted histogram labelled "$\pi$". (b) The cross sections of part (a) multiplied by $E^{-0.7}$; in the cases of photodisintegration into D, and disintegration of $^4$He during photoproduction, the curves have been multiplied by 10.

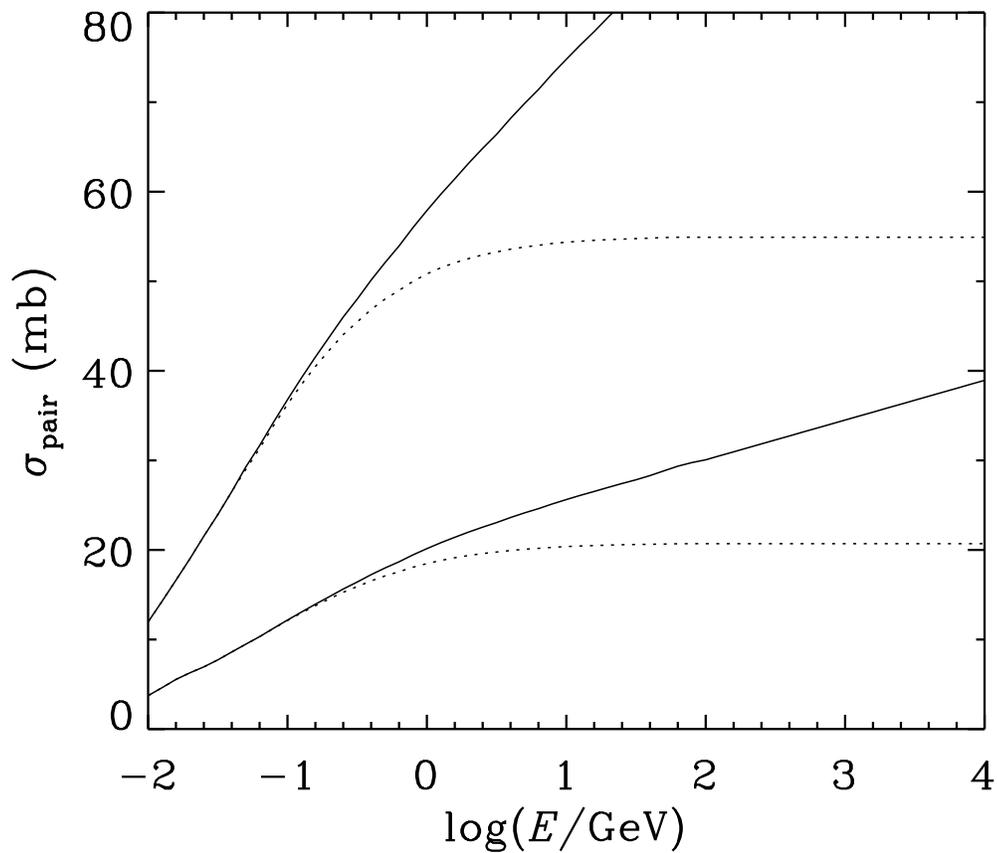

Figure 3: The pair production cross sections for hydrogen (lower curves) and helium (upper curves) for the two cases: fully ionized matter (full curves), and neutral matter (dotted curves).

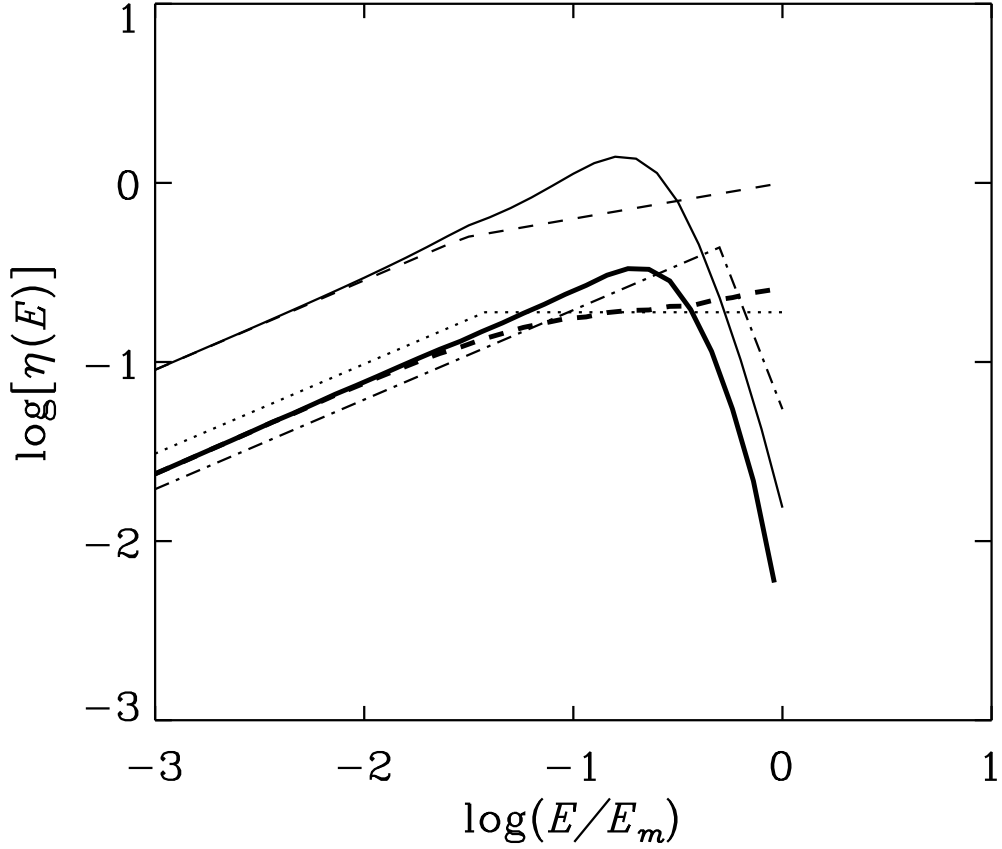

Figure 4: The escaping spectrum of photons from cascades on black body radaition for $E_m = 4E_c$ (see text) in which photon-photon scattering is included (solid curves) or neglected (dashed curves). The lower (thick) curves show results from the present Monte Carlo/Matrix calculation while the upper (thin) curves are from Svensson and Zdziarski [19]. Results of Svensson and Zdziarski are arbitrarily normalized such that the case for no scattering has $\eta(E_m) = 1$. Also shown are our approximate spectrum given in Eq. 17 (dotted curve), and the spectrum used in ref. [1] (chain curve). (All spectra cut off at $E = E_m$.)

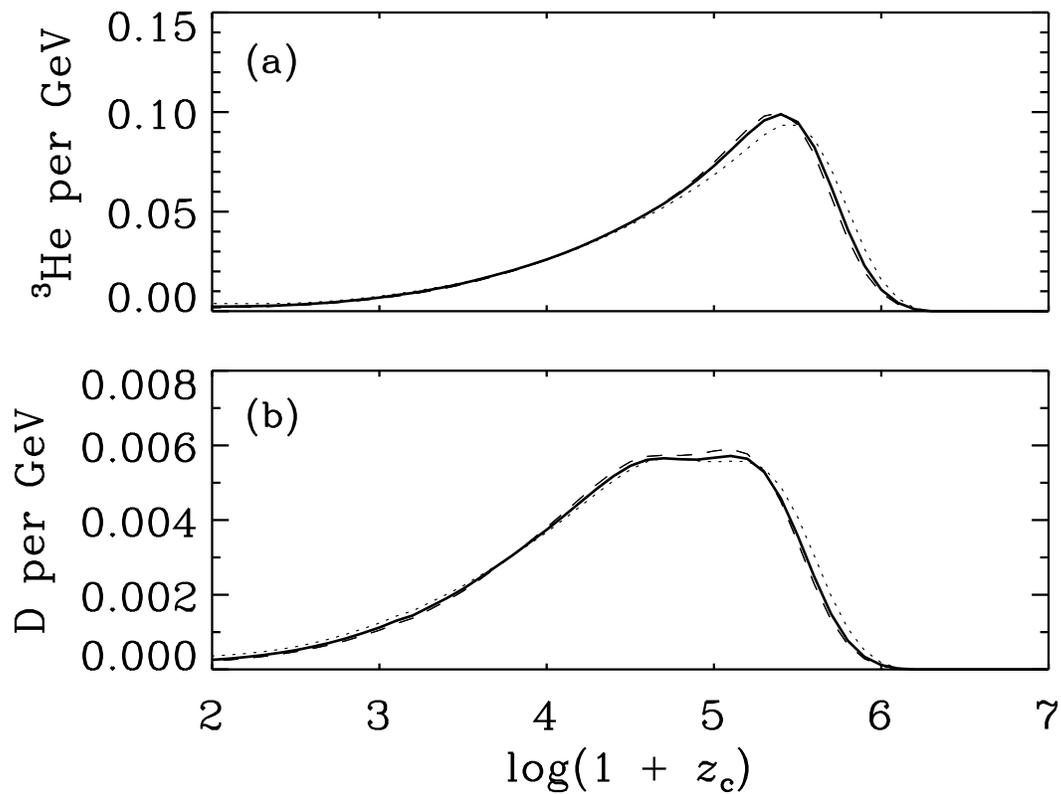

Figure 5: The number of (a) $^3$He nuclei and (b) D nuclei produced per GeV of cascade energy at redshift $z_c$. Results are shown for various $h$ (and $\Omega_b$): dotted curves – 0.4 (0.125); full curves – 0.7 (0.025); and dashed curves – 1.0 (0.01).

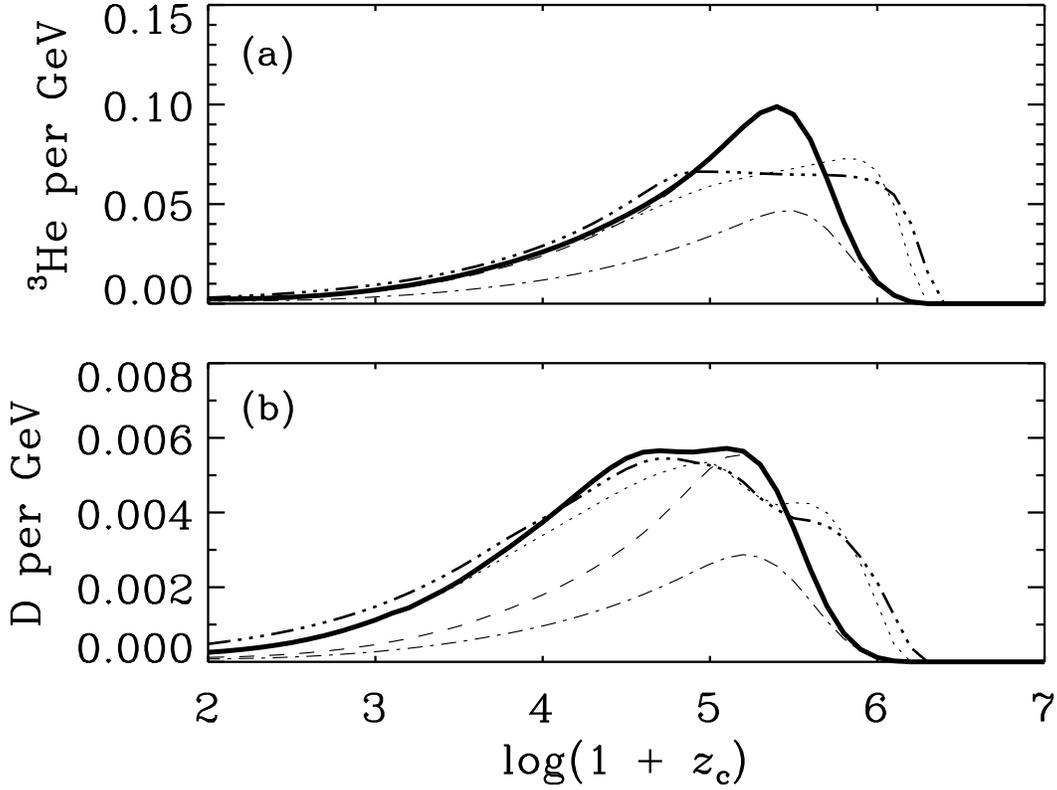

Figure 6: The number of (a) $^3$He nuclei and (b) D nuclei produced per GeV of cascade energy at redshift $z_c$. Results are shown for $h = 0.7$ and $\Omega_b = 0.025$ and different assumptions: ——— accurate calculation including all effects; ·········· photon-photon scattering neglected; — — — — disintegration of $^4$He during pion photoproduction neglected; — · — · — · — · — asymptotic pair production and bremsstrahlung cross sections used, and disintegration of $^4$He during pion photoproduction neglected; — · · · — · · · — approximate treatment.

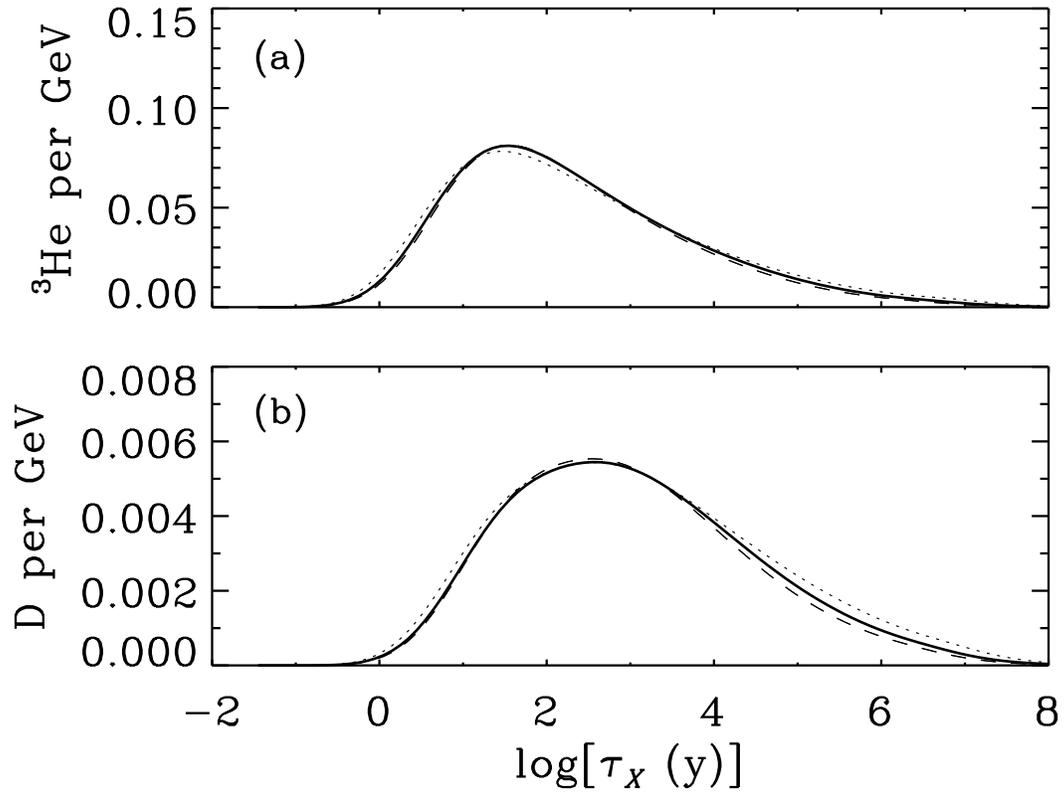

Figure 7: The number of (a) $^3$He nuclei and (b) D nuclei produced per GeV of cascade energy as a function of mean decay time. Results are shown for various $h$ (and $\Omega_b$): dotted curves – 0.4 (0.125); full curves – 0.7 (0.025); and dashed curves – 1.0 (0.01).

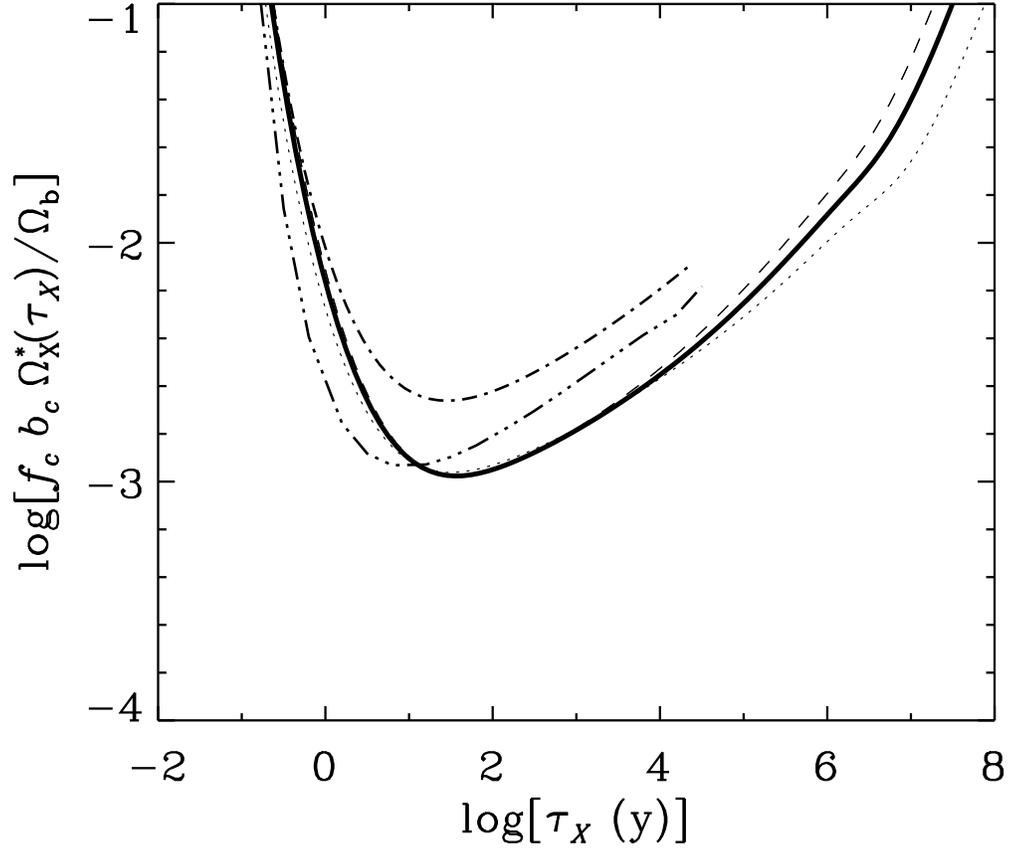

Figure 8: Upper limit, $\Omega_X^*$, to the fraction of the present closure density that massive particles would contribute if they had not decayed with with mean decay time $\tau_X$, multiplied by $f_c b_c/\Omega_b$. Results are shown for various $h$ (and $\Omega_b$): dotted curves – 0.4 (0.125); full curves – 0.7 (0.025); and dashed curves – 1.0 (0.01). Also shown are results we would obtain if we used asymptotic pair production and bremsstrahlung cross sections (— · — · — · — ·—), and results of Ellis *et al.* [19] (— ··· — ··· —).

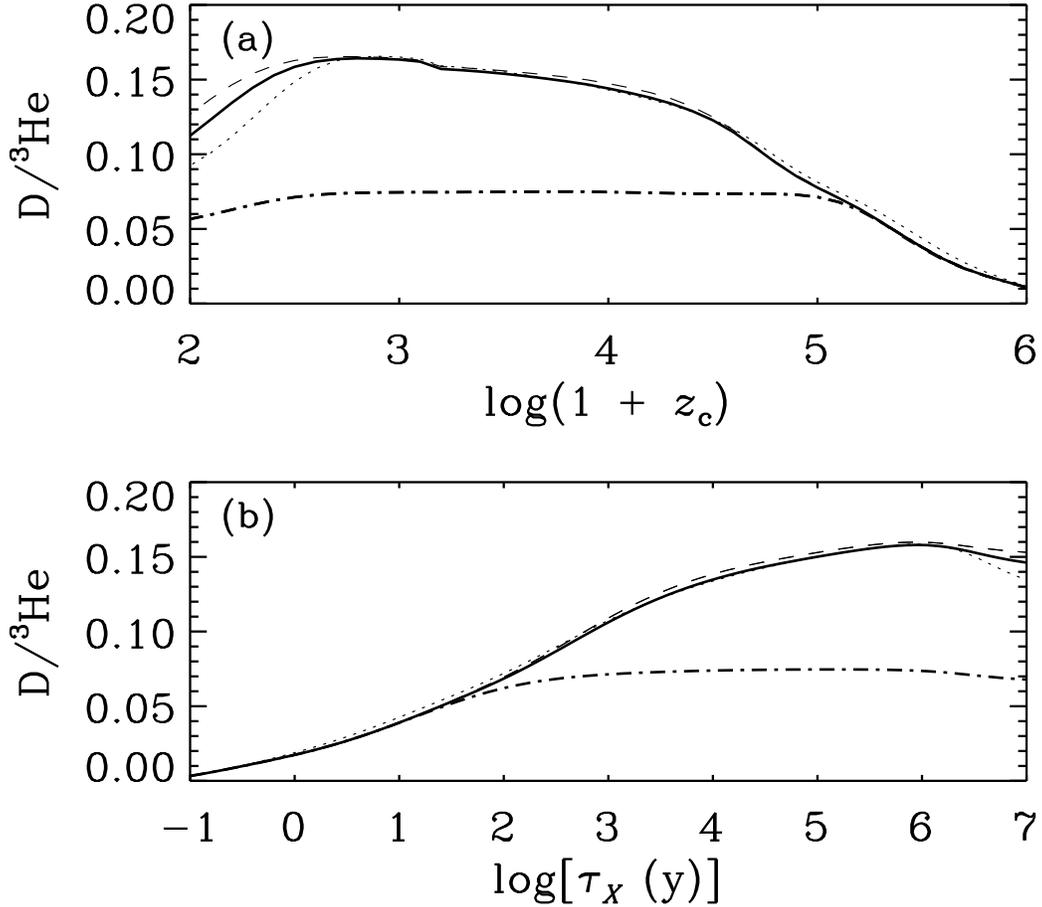

Figure 9: Ratio of D to $^3$He for massive particle decay as a function of: (a) redshift at decay, $z_c$; (b) mean decay time, $\tau_X$. Results are shown for various $h$ (and $\Omega_b$): dotted curves – 0.4 (0.125); full curves – 0.7 (0.025); and dashed curves – 1.0 (0.01). The dot-dash curve shows the result for $h = 0.7$ and $\Omega_b = 0.025$ that would be obtained if disintegration during photoproduction were neglected.